# Gas production from methane hydrates upon thermal stimulation; an analytical study employing radial coordinates

M. Roostaie[1] and Y. Leonenko[1,2,*]

[1]*Department of Earth & Environmental Sciences*

[2]*Department of Geography and Environmental Management*

*University of Waterloo*
*200 University Ave. W*
*Waterloo, Ontario, Canada N2L 3G1*

*corresponding author, leonenko@uwaterloo.ca, 519-888-4567 (Ex. 32160)

**Abstract**

In this study, a radial analytical model for methane hydrate dissociation upon thermal stimulation in porous media considering the wellbore structure's effect has been developed. The analytical approach is based on a similarity solution employing a moving boundary separating the dissociated and undissociated zones. Two different heat sources are considered: i) line heat-source; and ii) wellbore heat-source with a specific thickness consisting of casing, gravel, and cement. The temperature and pressure distributions, dissociation rate, and energy efficiency considering various reservoir properties and different initial and boundary conditions are investigated. Direct heat transfer from the heat source to the reservoir without considering heat conduction in the wellbore thickness causes a higher dissociation rate and gas production in the line-heat-source model compared to those of the wellbore-heat-source model. Increasing the heat-source temperature or decreasing its pressure increases gas production. However, employing them simultaneously results in greater gas production but reduces energy efficiency. The dissociation rate has direct relation with reservoir's porosity, thermal diffusivities, and thermal conductivities, but it is not dependent on the reservoir's permeability.

**Introduction**

Gas hydrates are ice-like components known as clathrates [1]. These consist of gas molecules, such as methane, trapped inside crystalline water molecules which can be formed wherever a sufficient amount of gas and water in combination with high pressure and low temperature conditions exist [2]. Collett [3] and Chong et al. [4], respectively, through an assessment on the hydrate reservoirs in North America and a review on gas production from hydrates reported that such conditions can be found in deep permafrost regions and submarine zones. One volume of methane hydrate (MH) can produce approximately 164.6 volumes of methane and 0.87 volumes of water at standard pressure and temperature conditions [5]. MH can make a significant contribution to future sources of energy as the estimate of the total volume of available MH exceeds the total volume of conventional gas resources worldwide [6-8]. However, more investigation is required to thoroughly understand the behavior of this complicated material and to make gas production from MH economically possible.

Considering that the entrapped gas and water are bonded through physical interaction in MH, which is weaker than the chemical interactions, any change in MH equilibrium temperature or pressure would easily induce MH instability and dissociation. The main methods of MH dissociation that have been reported so far are as follows [1, 9]: i) thermal stimulation by increasing the temperature of the reservoir above the equilibrium temperature. Li et al. [10], Fitzgerald et al. [11], and Yu et al. [12] respectively employed huff and puff, down-hole combustion, and hot water circulation methods as thermal stimulation; ii) depressurization by decreasing the pressure inside the MH reservoir below the equilibrium pressure of the hydrate. Yousif et al. [13] experimentally and mathematically investigated hydrate dissociation of Berea sandstone cores, located in USA upon depressurization. Yu et al. [14], and Feng et al. [15] numerically employed depressurization to produce gas from Nankai Trough, located in Japan. Ji et al. [16] and Terzariol et al. [17] respectively provided 2D and 1D analytical models to investigate hydrate dissociation upon depressurization; iii) depressurization in conjunction with thermal stimulation. Chong et al. [4, 17], Wang et al. [18], Jin et al. [19], and Wan et al. [20] respectively through a review, experimental, numerical, and experimental-numerical work reported that this method has a better efficiency compared to the two previous methods; iv) inhibitor injection by injecting fluids that will induce instability of the MH formation. For example, methanol and ethylene glycol are of the most popular inhibitors [4, 21] employed even in conjunction with depressurization [22]; and v) replacement of methane by $CO_2$ in MH reservoirs [23, 24]. The latter method is reported to help global warming and climate change mitigations [25-27]. For instance, Maruyama et al. [28] and Chen et al. [29] proposed power generation systems to dissociate MH beneath the sea floor and by the $CO_2$ capture and storage. Khlebnikov et al. [27] experimentally combined $CO_2$ replacement and inhibitor injection methods and received better results compared to using only one method. Despite several findings regarding the hydrate dissociation methods, further investigation is required to shed more light on the potential of different methods and their aspects.

Mathematical studies of hydrate dissociation can be categorized into: i) analytical approaches with fast answers and a better mechanistic understanding of the phenomena; and ii) numerical methods, which are more comprehensive and complicated, requiring fewer assumptions. In 1982, a 3D numerical model was developed to study MH dissociation [30]. In 1986, this work was extended by considering the effect of the water flow produced during the dissociation [31]. In 1991, a numerical model investigated MH dissociation by depressurization considering three phases of water, gas, and MH [32]. The previous work was then extended by taking into account convective-conductive heat transfer [33, 34]. MH dissociation upon thermal stimulation was simulated by assuming an impermeable moving dissociation boundary, which separates the dissociated and undissociated zones [35]. Another numerical work showed that dissociation upon depressurization is a function of well pressure [36]. In the same year, TOUGH2 simulator was employed to show the possibility of gas production from MH using both depressurization and thermal stimulation [37]. Results retrieved from TOUGH2 showed the feasibility of gas production from MH reservoirs in the Mackenzie Delta, located in Canada, upon depressurization and thermal stimulation with higher efficiencies by using both methods together [38]. Another numerical work reported that the kinetic reaction models should be taken into account in order to avoid under-prediction of recoverable MH, although requiring more computational effort compared to the equilibrium reaction models [39]. A numerical work using TOUGH-Fx/HYDRATE simulator showed low gas production with high amount of water production upon depressurization from disperse oceanic MH reservoirs with low hydrate saturation [40].

In 1990, Selim and Sloan [41] investigated MH dissociation upon thermal stimulation by using a 1D analytical model and assuming a moving dissociation boundary. In 1982, a study on hydrate dissociation upon both thermal stimulation using hot water circulation into the reservoir and

[Type here]

depressurization was performed [42]. Makogon [43] provided analytical expressions for the temperature and pressure distributions during MH dissociation upon depressurization by assuming a moving dissociation boundary. This work was extended by considering the water and gas movement [44]. In 2001, another model was generated based on Makogon's model [16] including heat conduction. An analytical work using depressurization reported that the effect of the gas-water two-phase flow on MH dissociation is smaller than the effect of heat transfer and the intrinsic kinetics of MH decomposition [45]. Recently, an analytical work based on experimental conditions (i.e., the reservoir was assumed to be finite, and there was heat transfer from outside of the reservoir into the hydrate zone) studied MH dissociation by depressurization, thermal stimulation, and the combination of both methods [46].

The experimental setup size is reported as one of the major challenges in MH dissociation investigations using different methods, which significantly affects the test outcomes [47]. For instance, this setup size determines the main mechanism in the hydrate dissociation [47], which involves any one of the following: i) heat transfer in the decomposing zone; ii) the intrinsic kinetics of hydrate decomposition; or iii) the multiphase flow (i.e., gas-water flow) during gas production [45]. Tang et al. [48] showed that the determining factor in the core-scale experiments is the intrinsic kinetics of hydrate decomposition; although in larger scale experiments or field works, the controlling mechanism is heat transfer in the decomposing zone. An experimental work in a 3D cubic hydrate simulator showed that MH dissociation using thermal stimulation is a moving boundary ablation process [49]. Li et al. [50] employed two hydrate simulators with different scales to experimentally investigate the MH dissociation upon depressurization. An experimental work showed that the main mechanism for heat transfer to the dissociating zone is conductive heat transfer [51]. Wang et al. [47] reported that ice formation in pores during MH dissociation below the quadruple point in the sandy sediment increases dissociation rate when employing a 3D Pilot-Scale Hydrate Simulator (PHS). Another experimental work that studied hydrate dissociation using PHS and reported that depressurization in conjunction with thermal stimulation is the optimum method [52].

Nowadays, this field of investigation has attracted researchers' interest to perform more mathematical studies as well as real field or experimental works employing parameters from real reservoirs [53]. Konno et al. [54] conducted the first offshore gas production test from MH reservoirs at the eastern Nankai Trough in Japan. Chen and Hartman [55] employed microfluidic systems with to investigate MH dissociation. Mardani et al. [56] experimentally investigated the effect of different kinetic hydrate inhibitors on MH stability. Wang et al. [57] studied the effect of different scales of experimental setups on MH dissociation. Zhao et al. [58] numerically investigated the effect of the reservoir's permeability on MH dissociation upon depressurization. Ding et al. [59] employed $CO_2/H_2$ gas mixture to replace methane in MH reservoirs using experimental approaches.

Studies over the past decades have provided substantial information about MH dissociation and the associated consequences in the reservoirs. However, it should be noted that the wellbore structure can play an important role in the dissociation process, such as the heat transfer mechanism during the thermal stimulation method. Recently, Roostaie and Leonenko [60] mathematically showed that the wellbore structure, consisting of multiple layers-casing, gravel, and cement- can also affect the interactions in the reservoir and the process efficiency upon wellbore heating. They designed an analytical semi-infinite 1D model and verified it against previous numerical and experimental works. Surprisingly, other analytical studies conducted on hydrate dissociation to present [58] have not considered the impact of wellbore geometry and the associated structure (i.e., wellbore radius and the associated external layers) on MH dissociation upon thermal stimulation by wellbore heating, which make them unreliable compared to experiments or field work. A systematic and detailed understanding of how aspects of the wellbore structure contribute to MH dissociation is still lacking. It should also be noted that employing radial coordinates in the previous analytical studies of thermal stimulation have not been treated in great detail. Thus, this study aims to obtain data which will help to address these research gaps and highlights the wellbore structure's importance in the dissociation process by expanding the previous work [60] employing an infinite 2D radial geometry and cylindrical wellbore with external layers along with another model with a line heat-source. These assumptions render the outcomes closer to the real operational and practical conditions.

The present work develops 2D analytical models assuming an infinite hydrate reservoir in radial coordinates and two different heat sources: i) line heat-source (no thickness); and ii) wellbore heat-source consisting of three main completion layers of casing, cement, and gravel. The energy efficiency, gas production, and temperature and pressure distributions are calculated and verified against the previous experimental and mathematical studies. The results of this work shed more light on assessing the gas production from MH reservoirs upon thermal stimulation. Therefore, this study makes a major contribution to research on MH dissociation upon wellbore heating by analytically demonstrating, for the first time, the effect of wellbore geometry and structure on the MH dissociation in radial coordinates. The outcomes obtained using such conditions are closer to the real-condition tests, making them more valuable and reliable.

- **Modeling**

Figure 1 shows a schematic of the hydrate dissociation in the proposed 2D radial geometry. The dashed circle shows the moving dissociation boundary, and the grey region denotes the wellbore thickness consisting of cement, gravel, and two layers of casing. There are many different wellbore structures and geometries assumed in the literature for different purposes, such as rock fracturing [61], SAGD steam injection [62], and oil recovery [63]. The proposed wellbore structure in this study is taken as a general model as understood to the best of our knowledge from the previous studies about gravel-packing techniques [64], sand-control purposes [65], wellbore's thermal resistance [66], and gas extraction from the MH reservoirs by depressurization [64, 67].

The geometry of the other case using a line heat-source is exactly the same as Figure 1, but it uses a line heat-source without thickness in the center of the reservoir instead of a wellbore with a specific thickness. The following steps are assumed as the basics of MH dissociation: i) before the dissociation begins, the reservoir has a porosity of $\phi$ and is filled with MH in equilibrium conditions of temperature $T_0$; ii) at time $t=0$, the heat source warms up by increasing its temperature (the inner surface of the heat source for the wellbore-heat-source case) to a new temperature $T_i$, which is higher than the hydrate equilibrium temperature, and is kept constant afterward; and iii) MH dissociation begins with a sharp moving boundary surface showing the rate of hydrate dissociation and separating the water and gas produced in the dissociated zone (Zone I) from the undissociated zone (Zone II).

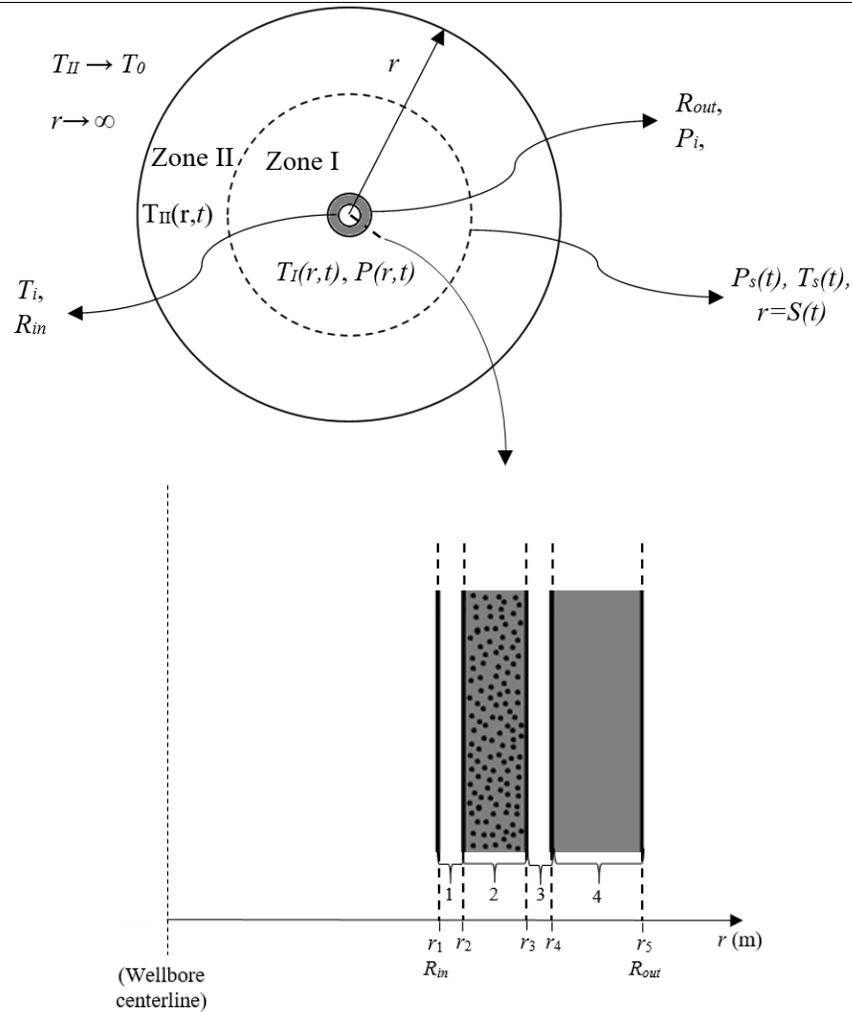

Figure 1. Schematic of hydrate dissociation model in infinite radial model. The dissociation interface is identified by the dashed circle. The grey region shows the wellbore thickness. Different parts of the wellbore structure are: 1) casing I, 2) gravel pack, 3) casing II, and 4) cement

During dissociation, the temperature of Zone I is higher than that of Zone II inducing heat transfer from the Zone I to Zone II. Principally, heat input from the heat source is consumed in two different ways: i) the temperature increment of the matrix sediments and the produced water and gas in Zone I; and ii) hydrate dissociation and the temperature increment of the matrix materials in Zone II. Over time, Zone I becomes larger and absorbs higher amount of input heat (the first way mentioned above) reducing the dissociation rate and the moving interface speed.

Figure 2 shows the temperature and pressure distributions in the system after dissociation begins. The temperature and pressure distributions in Zone I ($T_I$ and $P$) are respectively $T_s < T_I < T_i$ and $P_i < P < P_s$, and the temperature distribution in Zone II ($T_{II}$) is $T_0 < T_{II} < T_s$ by assuming a constant hydrate pressure equal to the equilibrium pressure in this zone. It should be noted that $T_I$ at the outer surface of the well changes over time, but it is always lower than $T_i$.

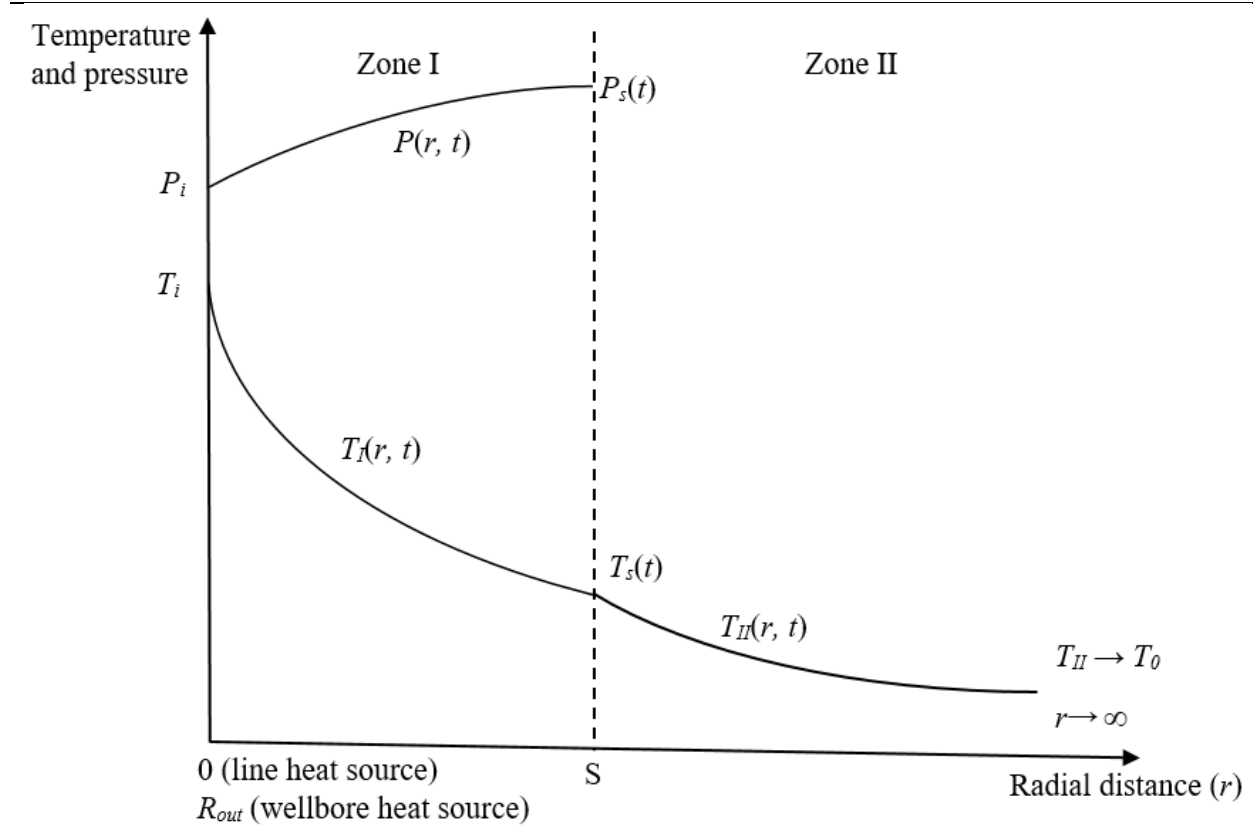

Figure 2. Schematic of pressure and temperature distribution in the reservoir upon hydrate dissociation.

The produced gas will be streaming towards the heat source according to Darcy's Law and induces a sudden change in density at the dissociation front. Other assumptions made in the models of this study, which are consistent with the previous analytical works [41, 60, 68, 69], are: i) thermodynamic equilibrium at the dissociation interface for temperature and pressure; ii) the water produced from the dissociation process remains motionless in the pores of Zone I; iii) constant thermophysical properties of the phases; iv) the produced gas shows an ideal behavior; v) the produced gas instantaneously reaches thermal equilibrium with the local sediments; vi) no viscous dissipation or inertial effects.

[Type here]

Basic equations which are the same for both types of heat sources (line heat-source and wellbore heat-source), are presented in the following formulas.

The continuity equation of gas in Zone I is:

$$\phi\left(\frac{\partial \rho_g}{\partial t}\right) + \left(\frac{\partial \rho_g v_g}{\partial r}\right) = 0, \ t > 0 \tag{1}$$

where $\phi$ is reservoir (matrix) porosity, $\rho_g$ is gas density (kg/m³), $v_g$ is gas velocity (m/s), and $r$ is the radial distance (m). The gas velocity in Zone I is calculated by using Darcy's Law:

$$v_g = -\left(\frac{k}{\mu}\right)\left(\frac{\partial p}{\partial r}\right), \ t > 0 \tag{2}$$

where $k$ is gas permeability (m²), $\mu$ is gas viscosity (mPa.s), and $P$ is pressure (Pa).

Equations 3 and 4 show the energy balance in Zones I and II, respectively:

$$\rho_I C_{pI} \frac{\partial T_I}{\partial t} + \frac{\partial \rho_g C_{pg} v_g T_I}{\partial r} = k_1 \frac{\partial}{\partial r} r \frac{\partial T_I}{\partial r}, \ t > 0, \ \text{in Zone I} \tag{3}$$

$$\frac{\partial T_{II}}{\partial t} = \frac{\alpha_{II}}{r} \frac{\partial}{\partial r} r \frac{\partial T_{II}}{\partial r}, \ t > 0, \ \text{in Zone II} \tag{4}$$

where $\rho_I$ is the matrix density (kg/m³) in Zone I, $\rho_g$ is gas density (kg/m³), $C_{pI}$ is the matrix specific heat capacity (J/(kg.K)) in Zone I, $C_{pg}$ is the gas specific heat capacity (J/(kg.K)), $T_I$ is the matrix temperature (K) in Zone I, $T_{II}$ is the matrix temperature (K) in Zone II, $\alpha_{II}$ is the matrix thermal diffusivity (m²/s) in Zone II, and $k_I$ is the matrix thermal conductivity (W/(m.K)) in Zone I.

The following equation evaluates the gas density in Zone I by using Ideal Gas Law:

$$\rho_g = \frac{mP}{RT_I}, \ R_{out} < r < S, t > 0 \tag{5}$$

where $m$ is gas molecular mass (kg/mol), and $R$ is the universal gas constant (J/(mol.K)). The above equations represent the fundamental concept of the process. The heat sources' temperature and pressure are constant and equal to $T_i$ and $P_i$, respectively. Equations 6 and 7 also state these conditions:

[Type here]

$$T_I = T_i, \begin{cases} \text{Line heat} - \text{source: } r=0, t>0 \\ \text{Wellbore heat} - \text{source: } r=R_{in}, t>0 \end{cases} \quad (6)$$

$$P = P_i, \begin{cases} \text{Line heat} - \text{source: } r=0, t>0 \\ \text{Wellbore heat} - \text{source: } r=R_{out}, t>0 \end{cases} \quad (7)$$

Equation 8 states the heat transfer equation through the wellbore thickness:

$$-k_I A_w \frac{\partial T_I}{\partial r} = \frac{(T_i - T_I)}{R_w}, r=R_{out}, t>0 \quad (8)$$

$$R_w = \frac{\ln(r_2/r_1)}{2\pi k_s} + \frac{\ln(r_3/r_2)}{2\pi k_g} + \frac{\ln(r_4/r_3)}{2\pi k_s} + \frac{\ln(r_5/r_4)}{2\pi k_c} \quad (9)$$

where $A_w$ is the wellbore area (m²), $R_w$ is the wellbore thermal resistivity (W/K), $k_x$ is the thermal conductivity (J/(s.m.K)), $r$ is radius (m), the $s$, $c$, and $g$ subscripts respectively stand for steel (casing), cement, and gravel, and the subscripts 1-5 are schematically shown on Figure 1.

The interface pressure is calculated by the Antoine Equation (equation 10), which is a thermodynamic relationship between the interface temperature and pressure:

$$P_s = \exp(A_a - B_a/T_s), r=S, t>0 \quad (10)$$

where $P_s$ is the pressure (Pa) and $T_s$ is the temperature (K) at the moving interface, and $A_a$ and $B_a$ are constants. Equations 11 and 12 respectively show the mass and energy balances at the dissociation interface. Equations 13-16 respectively represent the MH dissociation heat and the associated boundary conditions [41],:

$$F_{gH} \rho_H \left(\frac{dS}{dt}\right) + \rho_g v_g = 0, r=S, t>0 \quad (11)$$

$$k_{II} \frac{\partial T_{II}}{\partial r} - k_I \frac{\partial T_I}{\partial r} = \phi \rho_H Q_{Hd} \frac{dS}{dt}, r=S, t>0 \quad (12)$$

$$Q_{hyd} = c + dT_s \quad (13)$$

$$T_I = T_{II} = T_s(t), r=S, t>0 \quad (14)$$

$$T_{II} = T_0, r \to \infty, t>0 \quad (15)$$

[Type here]

$$T_{II} = T_0, 0 < r < \infty, t = 0 \tag{16}$$

where $\rho_H$ is the hydrate density (kg/m³), $k_{II}$ is the thermal conductivity (W/(m.K)) of Zone II, $Q_{Hd}$ is MH dissociation heat (J/kg), and $c$ and $d$ are constants. In equation 11, $F_{gH}$ is a constant that represents the methane gas mass ratio inside the hydrate (0.1265 kg CH$_4$/kg hydrate) [41].

The following equations 17-19 are obtained respectively from equations 1, 3, and 11 by employing equations 2 and 5 to eliminate the gas velocity and density.

$$\phi \frac{\partial}{\partial t}\left(\frac{P}{T_I}\right) - \frac{k}{\mu}\frac{\partial}{\partial r}\left(\frac{P}{T_I}\frac{\partial P}{\partial r}\right) = 0 \tag{17}$$

$$\rho_I C_{pI} \frac{\partial T_I}{\partial t} + \frac{kmC_{pg}}{\mu R}\frac{\partial}{\partial r}\left(P\frac{\partial P}{\partial r}\right) = \frac{k_1}{r}\frac{\partial}{\partial r} r \frac{\partial T_I}{\partial r}, \ t > 0, \text{ in Zone I} \tag{18}$$

$$F_{gH}\phi\sigma_H\left(\frac{dS}{dt}\right) - \frac{kmP}{\mu RT_I}\frac{\partial P}{\partial r} = 0, \ r = S, t > 0 \tag{19}$$

The similarity solution using a dimensionless parameter (equation 20) is employed for transformation, simplification, and solution of the equations mentioned above. This method, which was first introduced by Neumann [70, 71], satisfies the initial and boundary conditions and assumes the movement of the dissociation interface to be inversely proportional with the square root of time ($t^{1/2}$) as follows:

$$\lambda = \frac{r}{\sqrt{4\alpha_{II}t}} \tag{20}$$

On the moving dissociation interface, equation 20 becomes:

$$\beta = \frac{S}{\sqrt{4\alpha_{II}t}}, \ r = S, t > 0 \tag{21}$$

And, on the outer surface of the wellbore:

$$\lambda_{os} = \frac{R_{out}}{\sqrt{4\alpha_{II}t}}, \ r = R_{out}, t > 0 \tag{22}$$

The abovementioned equations are transformed by employing equations 20-22 and presented in the supplementary information.

[Type here]

By considering the procedure recommended by previous works [70, 71], the following solutions for the temperature distributions in the two heat source cases are assumed implementing the exponential integral (Ei) function:

Line heat-source: $\begin{cases} T_I(\lambda) = T_i - A\,Ei(-(a\lambda+b)^2) + A\,Ei(-b^2) \\ T_{II}(\lambda) = T_0 + B\,Ei(-\lambda^2) \end{cases}$ (23)

Wellbore heat-source: $\begin{cases} T_I(\lambda) = -A_1\,Ei(-(a\lambda+b)^2) + A_1\,Ei(-b^2) + C \\ T_{II}(\lambda) = T_0 + B_1\,Ei(-\lambda^2) \end{cases}$ (24)

$A$, $B$, $C$, $A_1$, $B_1$, $a$, and $b$ constants are defined in the supplementary information.

The pressure distribution in Zone I for both heat sources can be calculated from the equation S7 as follows:

$$P(\lambda) = \left( P_0^2 + \frac{4F_{gH}\phi\rho_H \alpha_{II} \mu R \beta}{km} \int T_I d\lambda \right)^{1/2}$$ (25)

Then, equations 26 and 27 show the pressure distributions for both heat sources by replacing and integrating $T_I$ from equations 23 and 24:

Line heat-source: $P = \left( P_0^2 + L(\beta)M(\beta)\lambda + AL(\beta)(N(0) - N(\lambda)) \right)^{1/2}$ (26)

Wellbore heat-source: $P = \left( P_0^2 + L(\beta)(K(\beta)\lambda - A_1 N(\lambda) - (K(\beta)\lambda_{os} - A_1 N(\lambda_{os}))) \right)^{1/2}$ (27) where $L(\beta)$, $M(\beta)$, $N(\lambda)$, and $K(\beta)$ are defined in the supplementary information. The obtained solutions for temperature and pressure distributions satisfy the basic equations and boundary conditions (equations 1-19) by direct substitution.

The following formula calculates the heat flux from the wellbore (J/(s.m$^2$)) as a function of time:

$$u_r = -k_I \frac{\partial T_I}{\partial r}, \quad r = R_{out}$$ (28)

Equation 28 can be transformed to an equal equation according to equation 20 (provided in the supplementary information).

[Type here]

Integrating equation 28 gives the total heat input into the reservoir from the heat source (J/m²) up to time $t$ as follows:

$$Q_{rt} = -k_I \int_0^t \frac{\partial T_I}{\partial r} dt, \quad r = R_{out} \tag{29}$$

The total volume of produced gas at standard temperature and pressure (STP) of dry gas can be calculated as follows:

$$V_{rp} = \frac{n_{rt} R T_{STP}}{P_{STP}} \tag{30}$$

where $V_{rp}$ and $n_{rt}$ respectively are the total volume (m³/m²) and total moles (mole/m²) of produced gas per average moving-interface surface area up to time $t$ at $T_{STP}$ and $P_{STP}$ as temperature and pressure of dry gas at STP conditions, respectively. Further details are provided in the supplementary information.

The ratio of the energy that could be produced from the produced gas combustion to the energy input from the heat source during hydrate dissociation defines the energy efficiency ratio. The following formula evaluates the dissociation energy efficiency [72]:

$$\eta_r = \frac{V_{rp} Q_g}{Q_{rt}} \tag{31}$$

where $\eta_f$ is the energy efficiency ratio, and $Q_g$ is the gas heating value (J/m³) at STP conditions.

- **Results**

As equation 20 indicates, $\beta$ represents the dissociation interface dimensionless position and velocity ($v_s / (4\alpha_{II} t)^{1/2}$). The value of $\beta$ only depends on $P_s$ and $T_s$ (equation S28), which depend on the heat source pressure and temperature (outer surface of the heat source in the wellbore-heat-source case) according to their associated equations in the previous section. The temperature at the wellbore outer surface is time-dependent, inducing variable $P_s$ and $T_s$ over time; however, the line heat-source temperature in the other case remains constant, inducing constant $P_s$ and $T_s$. The

[Type here]

proposed properties and parameters (Table 1) are based on the previous studies [41, 60, 73-75]. Figure 3 shows that $\beta$ increases at the beginning in the wellbore-heat-source case because the temperature at the wellbore outer surface increases over time, but tends to converge to the associated value of $\beta$ in the line-heat-source case as the temperature at the wellbore outer surface converges to the temperature inside of the well ($T_i$). The trend of $\beta$ is in accord with that of Roostaie and Leonenko [60].

Table 1. Parameters used in the modeling.

| Parameter | Value |
|---|---|
| Wellbore's inner surface radius, $R_{in}$ ($r_1$), m | 0.07 |
| $r_2$, m | 0.077 |
| $r_3$, m | 0.092 |
| $r_4$, m | 0.099 |
| Wellbore's outer surface radius, $R_{out}$ ($r_5$), m | 0.124 |
| Cement thermal conductivity, $k_c$, W/(m.K) | 0.933 |
| Gravel thermal conductivity, $k_g$, W/(m.K) | 0.4 |
| Casing (steel) thermal conductivity, $k_s$, W/(m.K) | 43.3 |
| Porosity, $\phi$ | 0.3 |
| Permeability, $k$, μm$^2$ | 1 |
| Zone I thermal diffusivity, $\alpha_I$, μm$^2$/s | 2.89×10$^6$ |
| Zone I thermal conductivity, $k_I$, W/(m.K) | 5.57 |
| Zone II thermal diffusivity, $\alpha_{II}$, μm$^2$/s | 6.97×10$^5$ |
| Zone II Thermal conductivity, $k_{II}$, W/(m.K) | 2.73 |
| Hydrate density, $\rho_H$, kg/m$^3$ | 913 |
| Hydrate dissociation heat, $Q_{Hd}$, J/kg | 446.12×10$^3$ −132.638$T_s$ |
| Gas heat capacity, $C_{pg}$, J/(kg.K) | 8766 |
| Gas viscosity, $\mu$, Pa.s | 10$^{-4}$ |
| Gas heating value at STP conditions, $Q_g$, MJ/m$^3$ | 37.6 |
| Methane molecular mass, $m$, g/mol | 16.04 |
| Methane gas mass ratio trapped inside the hydrate, $F_{gH}$ | 0.1265 |
| Universal gas constant, $R$, J/(mol.K) | 8.314 |

[Type here]

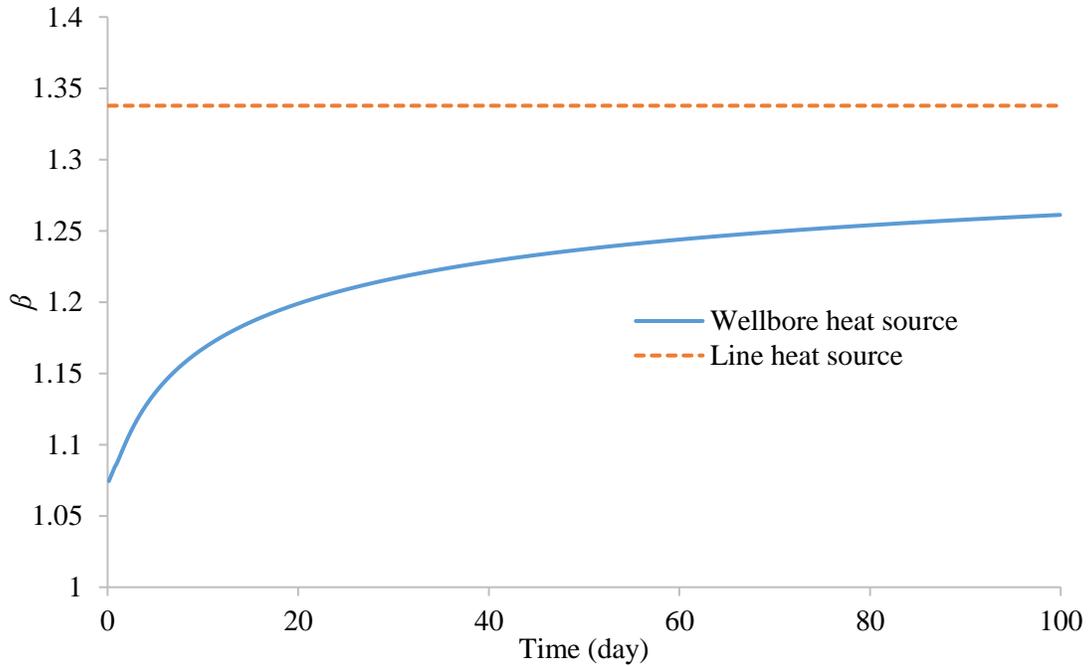

Figure 3. Dimensionless interface position assuming $T_0$=275 K $T_i = 563.5$ K, and $P_i = 7.6$ MPa.

Figure 4 represents the boundary conditions effect on the interface movement 100 days after the process begun using both heat sources. Increasing the pressure and decreasing the heat source temperature decrease $\beta$. The initial temperature of MH also has a direct impact on $\beta$. Previous studies have reported the same relation between $\beta$ and $P_i$, $T_i$, and $T_0$ [41, 60]. Another numerical work on MH dissociation upon depressurization, which was validated against Masuda's results [33], showed that decreasing wellbore pressure and increasing reservoir temperature increase the dissociation rate [76].

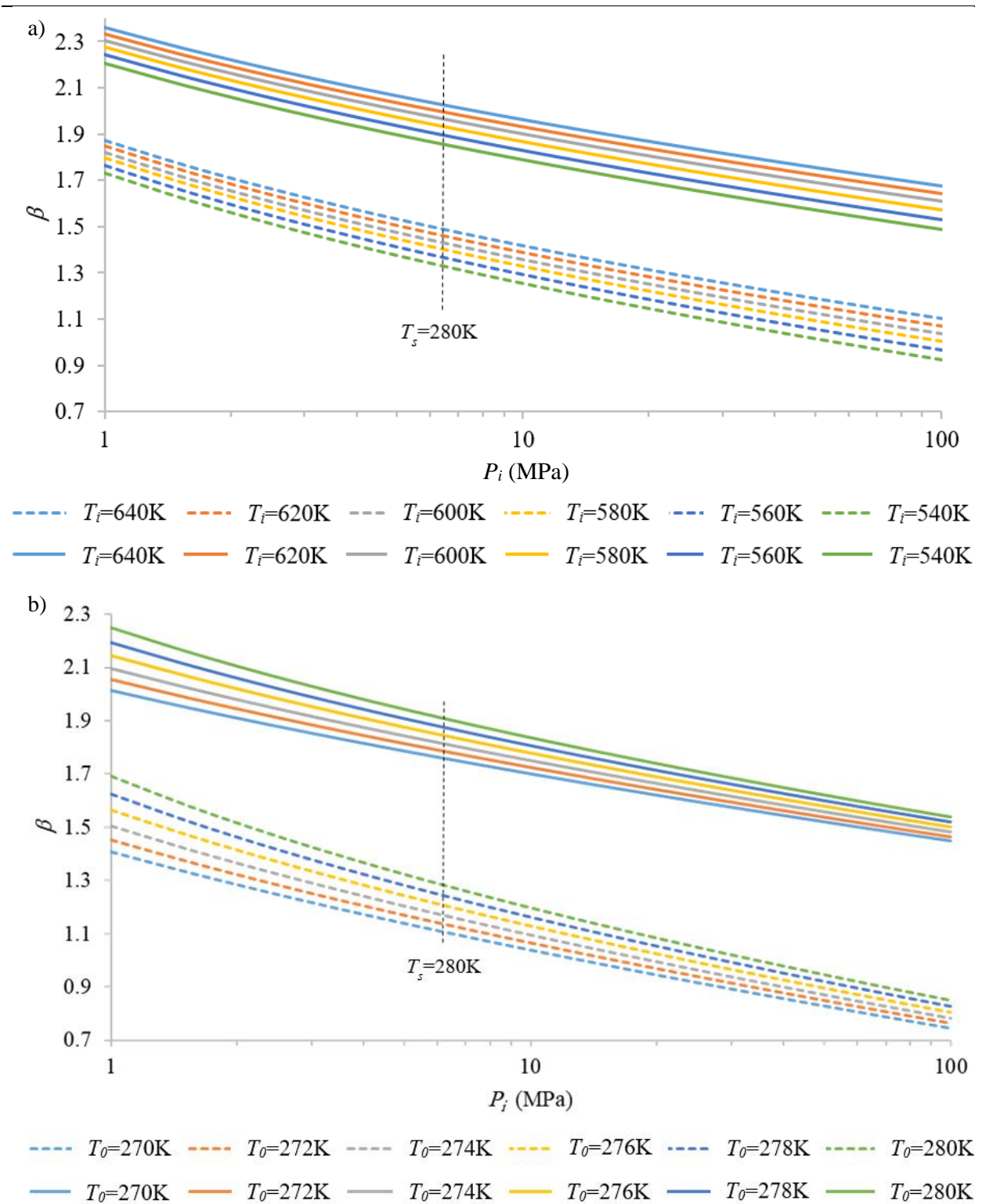

Figure 4. Dimensionless interface position assuming: a) $T_0$=280 K and various $T_i$ and $P_i$ values, and b) $T_i$=563.5 K and various $T_0$ and $P_i$ values. Dashed lines and solid lines respectively represent the wellbore-heat-source and the line-heat-source cases.

[Type here]

The vertical dashed lines in Figure 4 show the associated temperature and pressure at the locus on which the dissociation temperature equals 280 K. On the loci to the left (lower $P_i$) and right (higher $P_i$) of this locus, the dissociation temperature decreases and increases, respectively. This temperature is mainly dependent on heat source pressure, and almost independent of the heat source and MH temperatures. Previous works [41, 60] also reported that under lower heat source pressures of 6 MPa, $T_s$ may reduce to the freezing temperature of water, and ice generation can cease dissociation. For situations in which $T_s$ is higher than the MH temperature, some heat from the heat source will be consumed to increase the temperature of MH close to the dissociation front to $T_s$. If $T_s$ approaches $T_0$ all heat from the heat source is consumed for dissociation. Conversely, if $T_s$ falls below $T_0$, some part of the heat required for dissociation will be provided from the hydrate zone, reducing the temperature of this zone near the dissociation interface.

Temperature and pressure distributions for the two heat sources are calculated and presented in Figures S1 and S2 considering three different time frames and the following initial and boundary conditions (BCs): BC 1) $T_i = 450\,\text{K}$, $P_i = 10$ MPa, and $T_0 = 280\,\text{K}$, and BC 2) $T_i = 563.5\,\text{K}$, $P_i = 7.6$ MPa, and $T_0 = 275\,\text{K}$. The horizontal dashed lines in the temperature distribution diagrams (Figure S1) represent the temperature at the dissociation interface separating Zone I from Zone II. The trends for temperature and pressure distributions reported by Roostaie and Leonenko [60] and Selim and Sloan [41] are respectively similar to those of the cases with wellbore and line heat sources. Tsimpanogiannis and Lichtner [68], who built up a semi-analytical 1D model based on the work of Selim and Sloan [41] without the wellbore structure, showed that the higher temperatures of the heat source increased the interface pressure. The results of temperature distribution are also consistent with an experimental work performed by Li et al. [49] on MH dissociation upon thermal stimulation. They reported that the dissociation progressed by a moving boundary interface. Figure 5 shows the volume of gas produced (m$^3$) under STP conditions in both cases in response to different heat sources, as well as the amount of input heat (MJ/m$^2$), and energy efficiency for the model with wellbore heating considering two BCs over 100 days.

Various researchers have experimentally investigated MH dissociation by hot water circulation in a reservoir. for instance, using this approach, Song et al. [72] experimentally achieved a similar trend for the energy efficiency of gas production from MH . Their reported energy efficiency of 18 and 40 is in good agreement with the present results. Wang et al. [18] also found the same energy efficiency and gas production trends, and Li et al. [77, 78] for a 5.8 L cubic reactor, observed that the energy efficiency of the process is approximately 20.6. Wang et al. [79] reported an energy efficiency between 6 and 20. Tang et al. [80] reported that increasing the hot water temprature and decreasing the pressure of the wellbore improve the energy ratio (the same as the energy efficiency) of MH dissociation.

Taking a different approach, Bayles et al. [81] analytically studied MH dissociation upon cyclic steam circulation into the reservoir, and reported the same energy efficiency trend, convergence between 4 to 9.6, and gas production over one year's dissociation. The slight difference between their results and those reported here is most likely due to the direct steam circulation into the reservoir and the cyclic pattern of their process.

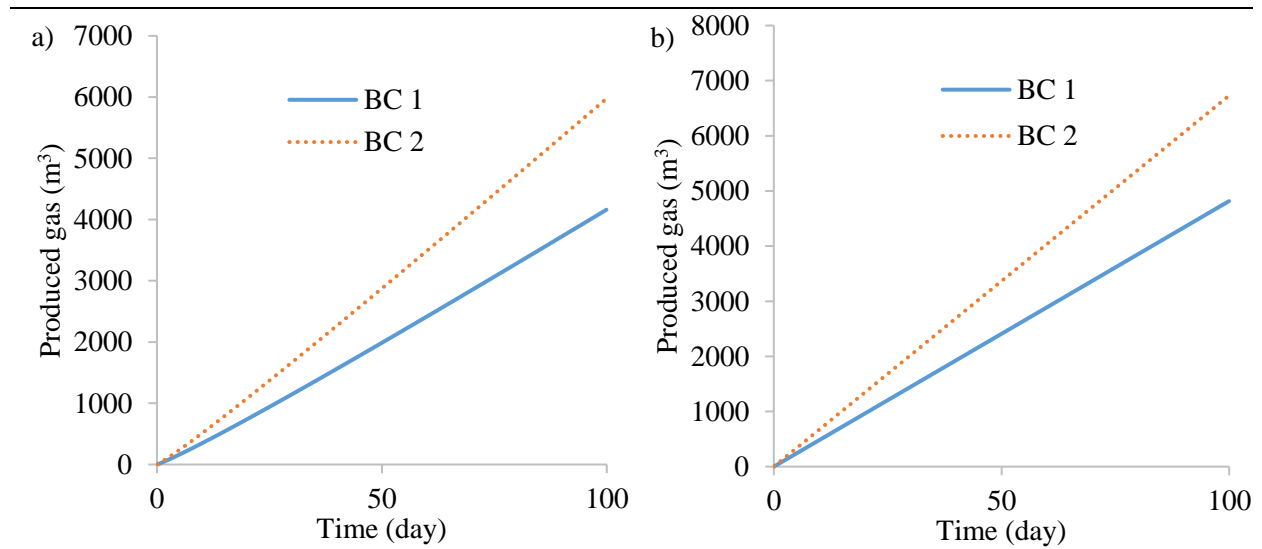

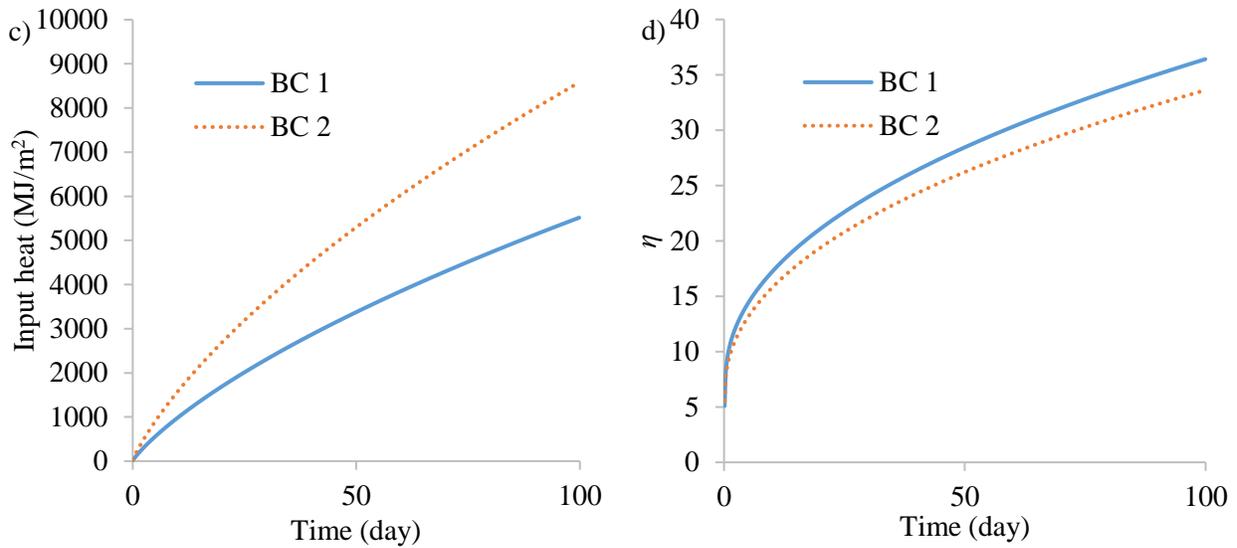

Figure 5. a) volume of gas produced in the wellbore-heat-source model, b) volume of gas produced in the line-heat-source model, c) amount of input heat in the wellbore-heat-source model, and d) energy efficiency during hydrate dissociation in the wellbore-heat-source model for two BCs.

A parametric study is employed to investigate the effect of various parameters (Table 2) on the dissociation process. Figures 6 and 7 respectively show the dissociation rate and gas-production resulting from the study. Figures S3 and S4 (supplementary file) respectively display the heat input from the wellbore to the reservoir and the energy efficiency in the wellbore-heat-source case. Increasing the thermal conductivity of Zone I significantly increases the amount of heat input from the reservoir (Figure S1a). Higher thermal diffusivity (lower heat capacity while the density is constant) causes the media to store less heat, which in turn, increases the heat transferred to the dissociation front. Eventually, higher thermal diffusivity and higher thermal conductivity increase the dissociation rate (Figure 6a) and gas production (Figure 7a). Lower thermal diffusivity in Zone II increases the storage of the heat transferred to this zone from the dissociation interface. Ultimately, this stored heat is released and consumed by dissociation, increasing the dissociation rate (Figure 6b). Higher thermal conductivity in Zone II reduces the dissociation rate and gas production because heat is being transferred to this zone from the moving interface more quickly than when it is consumed by dissociation. Higher thermal diffusivity in Zone II increases gas production (Figure 7b), as the reduction of heat storage in this zone induces more heat consumption by dissociation and more gas production. Lower thermal diffusivity and higher thermal

[Type here]

conductivity in Zone II increases the input heat (Figure S3b) for the same reason stated for input heat increment induced by the same changes in Zone I (Figure S3a). However, the input heat increment for Zone II is much lower than that of Zone I due to Zone I's direct contact with the wellbore. The energy efficiency in the case with wellbore heating has a direct relation to the thermal diffusivity of Zones I and II (Figures S4a and S4b). On the other hand, energy efficiency decreases with increasing thermal conductivity in Zones I and II. Higher porosity reduces the conduction heat transfer and the heat input from the wellbore (Figure S3c), ultimately reducing the dissociation rate (Figures 6c and 6d). Furthermore, higher porosity increases the amount of MH trapped in the pores eventually increasing the amount of gas produced (Figures 7c and 7d). Clearly, reservoir porosity has a direct relation on the energy efficiency of dissociation. Various permeabilities of media and gas viscosities have no impact on the dissociation process.

Selim and Sloan [41] and Roostaie and Leonenko [60] reported similar results for dissociation rate (Figure 6) during similar parametric studies. Zhao et al. [82] mathematically showed that increasing thermal conductivity had a direct positive effect on the dissociation process using hot water circulation in the reservoir; although water and gas relative permeabilities have almost no impact on the process. In another numerical work, they showed that increasing sediments' thermal conductivity increased the gas generation rate at the beginning of dissociation upon depressurization [83]. Both these works were verified against Masuda's experimental work [33]. Tsimpanogiannis and Lichtner [68] showed that increasing the porous media's thermal conductivity increased MH dissociation. Moridis et al. [38] conducted numerical analyses of various gas production scenarios from MH zones at the Mallik site and showed that a higher initial formation temperature, well temperature, and formation thermal conductivity increased gas production; although it is not affected by the formation's permeability and the specific heat of the rock and MH. It should be noted that the slight difference between the results of experimental works and the present work is due to different working conditions, such as direct hot water circulation into the reservoir, time period of experiments, and model parameters (i.e. hydrate saturation).

[Type here]

Table 2. Range of parameters employed in the parametric study.

| Parameter | Range |
|---|---|
| Porosity, $\phi$ | 0.1 to 0.5 |
| Permeability, $k$, μm² | 0.1 to 5 |
| Zone I thermal diffusivity, $\alpha_I$, μm²/s | $1\times10^6$ to $5\times10^6$ |
| Zone I thermal conductivity, $k_I$, W/(m.K) | 3 to 7 |
| Zone II thermal diffusivity, $\alpha_{II}$, μm²/s | $4\times10^5$ to $8\times10^5$ |
| Zone II thermal conductivity, $k_{II}$, W/(m.K) | 1 to 5 |
| Gas viscosity, $\mu$, Pa.s | $10^{-4}$ to $10^{-6}$ |

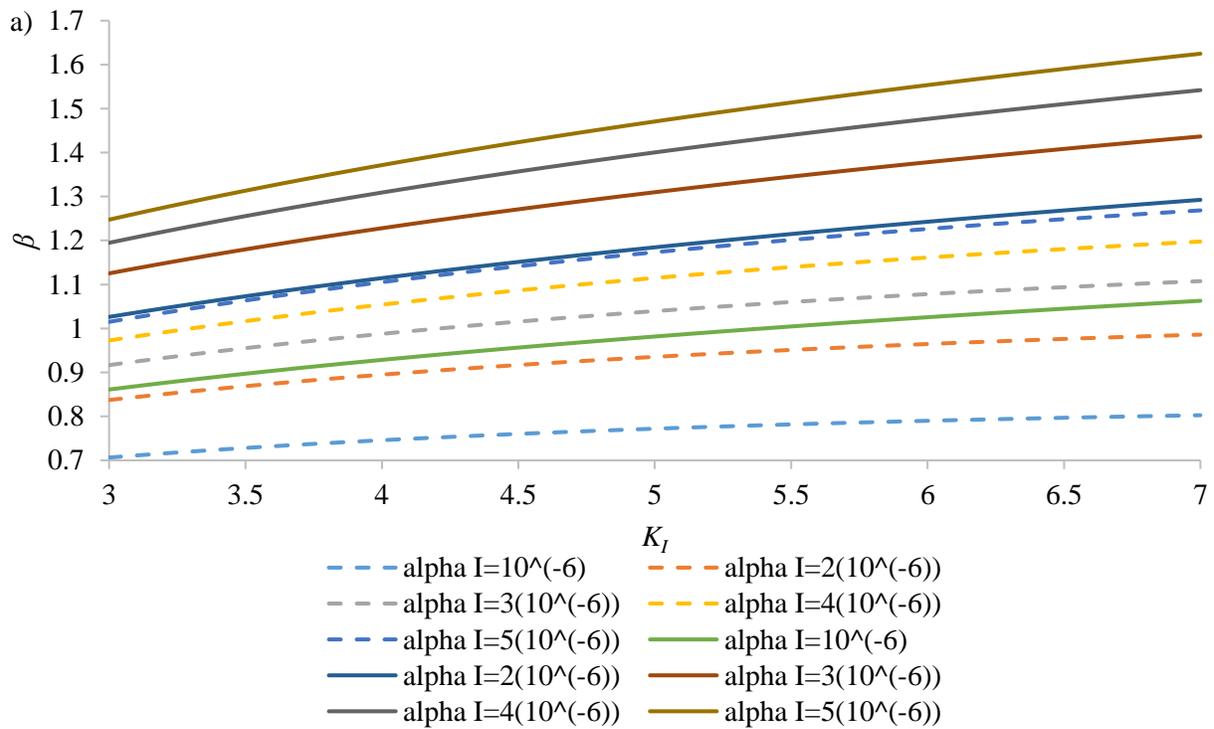

[Type here]

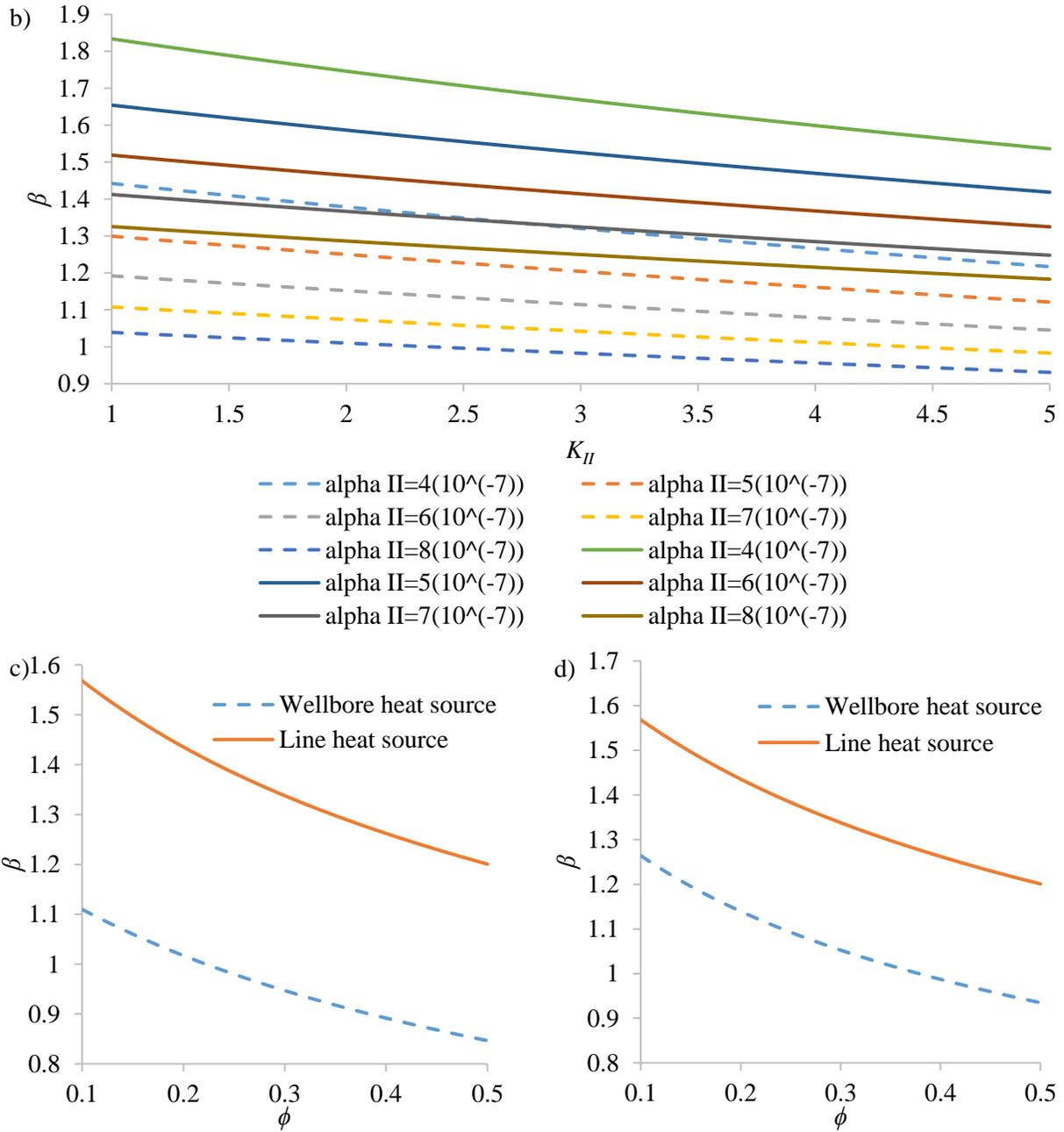

Figure 6. Effect of various parameters on the interface movement after 100 days dissociation considering both types of heat sources: a) thermal diffusivity and thermal conductivity of Zone I, b) thermal diffusivity and thermal conductivity of Zone II, c) porosity with various permeabilities, and d) porosity with various gas viscosities.

Dashed lines and solid lines respectively represent the wellbore-heat-source and the line-heat-source models.

[Type here]

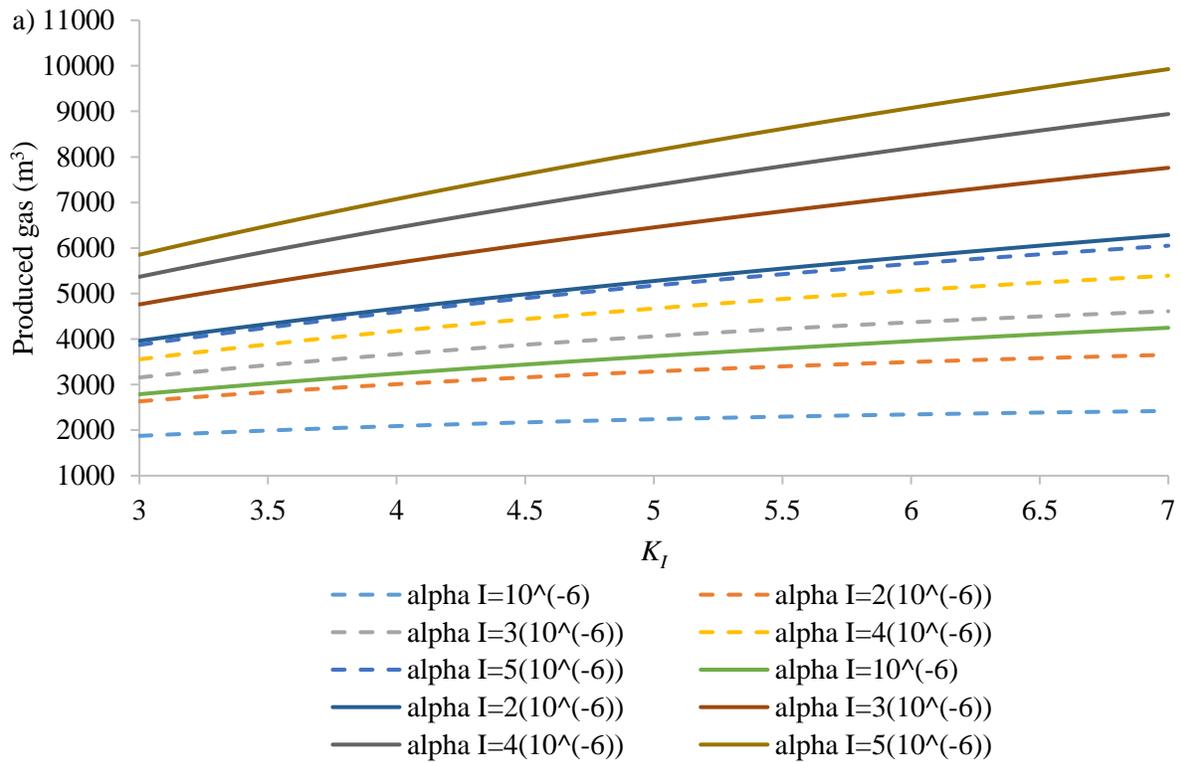

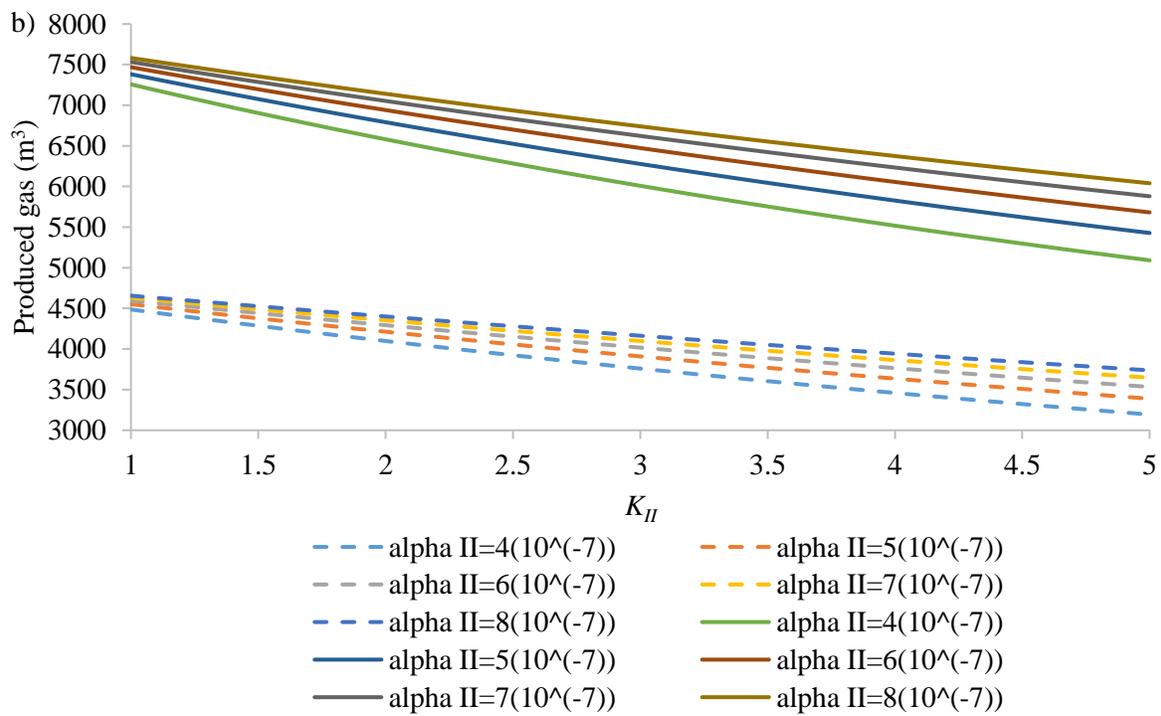

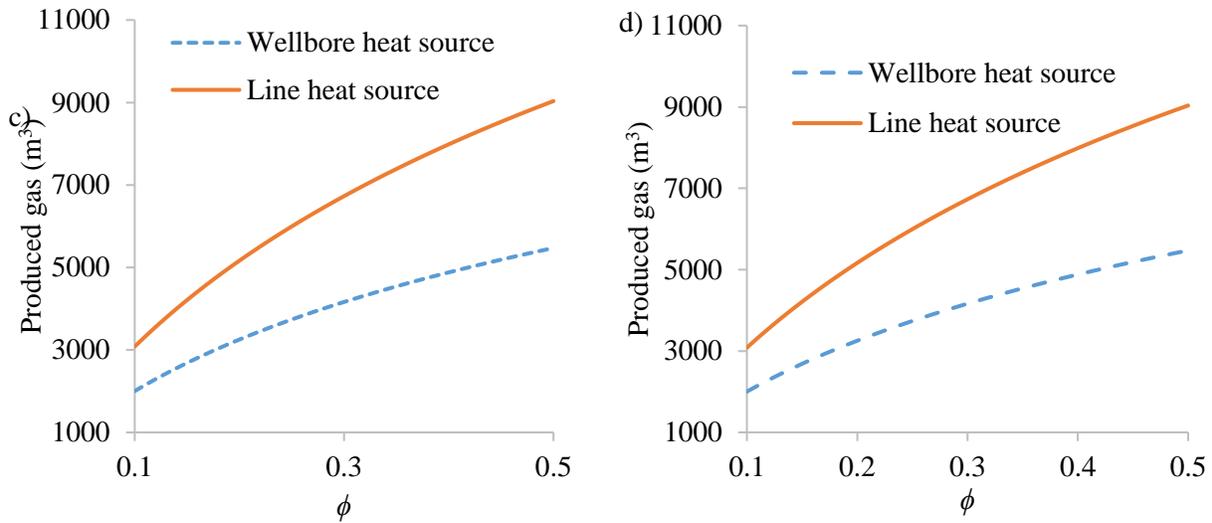

Figure 7. Produced gas after 100 days dissociation considering both heat sources and various parameters: a) thermal diffusivity and thermal conductivity of Zone I, b) thermal diffusivity and thermal conductivity of Zone II, c) porosity with various permeabilities, and d) porosity with various gas viscosities.

Dashed lines and solid lines respectively represent the wellbore-heat-source and the line-heat-source models.

- **Discussion**

The present study is a follow-up to our previous work on MH dissociation upon thermal stimulation [60], in which the wellbore structure effect on MH dissociation upon wellbore heating was assessed using a 1D analytical model in Cartesian coordinates. This 1D model is extended to 2D models by employing analytical approaches coupling the heat and mass transfer with MH dissociation in radial coordinates. Furthermore, two types of heat source are assumed: i) wellbore heat-source with a specific radius and external layers; and ii) line heat-source. Variable temperatures at the wellbore surface induced by the heat transfer in the external layers cause a lower dissociation rate, which is also inconstant, and produced gas compared to those of the line-heat-source case. Although the results reveal good agreement with the previous similar experimental and mathematical studies, none of the previous investigations analytically investigated the MH dissociation in radial coordinates upon wellbore heating by involving the effect of wellbore structure.

The magnitude of $\beta$ (Figure 3) is lower than those reported by Selim and Sloan [41] (approximately 32% lower than the line-heat-source case) and Roostaie and Leoeneko [60] (approximately 25% lower than the wellbore-heat-source case) due to the radial coordinate employed in this study. These results further prove that the dissociation rate is not constant and depends on the temperature at the wellbore surface, which changes by the conduction heat transfer in the wellbore structure.

The effect of the temperature increment of the wellbore's outer surface on the interface temperature becomes negligible as the dissociation progresses because over time; this temperature increment decreases (Figure S1) and Zone I becomes bigger, absorbing a larger part of the heat transferred from the wellbore. The distance between the dissociation interface and the heat source is longer in the line-heat-source model (Figures S1c and S1d) compared to that of the wellbore-heat-source model (Figures S1a and S1b). This is due to the direct heat transfer from the line heat-source to the reservoir that causes higher dissociation rates, while there is heat conduction in the wellbore thickness in the other model. This difference decreases over time as Zone I absorbs larger amount of input heat, which reduces the negative effect of heat source thickness on the dissociation rate. Interface pressure (Figures S2a and S2b) is not constant and increases due to the temperate increment at the wellbore surface but tends to converge to the associated values of the line-heat-source case (Figures S2c and S2d) as the temperature at the well surface approaches that inside the well. Selim and Sloan [41] and Roostaie and Leonenko [60] reported relatively smaller interface locations compared to those presented in Figures S1 and S2 due to the radial coordinates employed in the present study, inducing the interface area increment through the dissociation and absorption of higher amount of input heat by Zone I.

The gas produced is higher for the line-heat-source case compared to that of the wellbore-heat-source case due to the interface location mentioned earlier (Figures 5a and 5b). The gas produced and the input heat are higher by applying BC 2 compared to that of BC 1 (Figures 5a-5c). However, the energy efficiency is higher in the BC 1 case compared to that of BC 2 (Figure 5d) because the difference between the amount of input heat for the two BCs is higher than the difference between the associated amounts of produced gas. Thus, increasing the heat source temperature and decreasing its pressure increase the dissociation rate, but will not increase the efficiency of the process because Zone I becomes bigger and absorbs a larger part of the input heat which decreases the slope of produced gas and energy efficiency. Selim and Sloan [41] reported higher produced gas and energy efficiency by applying BC2 compared to those of BC1. They also showed a constant energy efficiency between 6.4-11.2. The heat transfer in the wellbore structure and the radial geometry of the present study caused these differences from the present results. Roostaie and Leonenko [60] showed similar trends for input heat and produced gas by applying the same initial and boundary conditions. They also reported higher energy efficiency for BC1 than that of BC2, but the associated trend of energy efficiency had a maximum peak in the beginning of the dissociation, then it decreased over time. As mentioned in their work, this behavior is due to the employed model geometry.

Lastly, the previous studies (mentioned in the previous section) mainly investigated effects of reservoir characteristics on the produced gas and dissociation rate without assessment of the associated effects on the input heat from the wellbore (Figure S3) and the energy efficiency (Figure S4), which would also clarify the results shown in Figures 6 and 7. The present results reveal that higher thermal conductivities of Zones I and II decrease the energy efficiency; although the produced gas is increased in higher Zones I's thermal conductivities. This is due to: i) the input heat increment induced by higher thermal conductivities (more pronounced for Zone I) (Figure S3a); and ii) the lower gas production caused by increasing Zone II's thermal conductivity. Increasing thermal diffusivities causes higher gas production and lower input heat resulting in higher energy efficiencies. Furthermore, the effect of thermal diffusivity of Zones I and II on the energy efficiency is more pronounced respectively in lower and higher thermal conductivities of Zones I and II. Higher thermal conductivities always increase the input heat, but higher porosity decreases it, inducing a significant raise in the energy efficiency.

- **Conclusions**

The present research aims to clarify how the wellbore structure affects the hydrate dissociation upon wellbore heating by employing 2D radial analytical models. Two types of heat sources are considered: i) line heat-source; and ii) wellbore heat-source with external layers (casing, cement, and gravel). The effects of various reservoir parameters and boundary conditions on dissociation are also evaluated. Dissociation interface position/velocity and energy efficiency are two main factors employed to assess the process. This project provided an important opportunity to advance the understanding of MH dissociation, as the analytical model introduced in this study appears to be the first work to investigate MH dissociation upon wellbore heating in radial coordinates, which is closer to the real conditions. Prior to this study it was difficult to make predictions about how the wellbore's properties and structure, through which heat transfer to the reservoir occurs, affect the dissociation. For example, the heat input and the resulted energy efficiency can be calculated, which are great assessments for the process. The results of the present study match with those of previous mathematical and experimental investigations. Taken together, the results make several contributions to the current literature as follows:

- The dissociation rate and process efficiency are dependent on the wellbore structure, which is a design parameter. Using a wellbore heat-source causes a reduction in the dissociation rate and the produced gas compared to those of the other case with line heat-source.
- Heat conduction through the wellbore's external layers changes wellbore surface temperature over time, which depends on the wellbore structure.
- Dissociation front's pressure and temperature depend on the wellbore surface temperature.
- Increasing the wellbore temperature and decreasing its pressure simultaneously increase the dissociation rate and produced gas while reduces the energy efficiency. Thus, more information on the initial and boundary conditions including wellbore structure (number of layers, thicknesses, and thermal properties) is essential to establish a greater degree of accuracy on this matter and to improve the energy efficiency of the process.
- Only conductive heat transfer was considered during the dissociation process in the proposed parametric study.
- Gas production, input heat, and energy efficiency of the process significantly depend on the reservoir's thermal diffusivity, porosity, and thermal conductivity. Inversely, different reservoir's permeabilities and viscosities do not affect the dissociation outcome.
- There is a good agreement between the present results and previous experimental and mathematical studies, validating the assumptions made in the model development.

[Type here]

## Acknowledgements

Financial support for this work provided by Natural Sciences and Engineering Research Council of Canada (NSERC)

| **Nomenclature** | | | |
|---|---|---|---|
| $A$ | Dimensionless variable in equation S9 | $P$ | Pressure in Zone I |
| $A_1$ | Dimensionless variable in equation S10 | $P_s$ | Interface pressure |
| $A_a$ | Dimensionless constant in equation 10 | $P_i$ | Heat source pressure |
| $A_s$ | Dissociation front average area | $P_{STP}$ | Gas pressure at STP conditions |
| $A_w$ | Wellbore area | $Q_{Hd}$ | MH dissociation heat |
| $a$ | Dimensionless constant in equation S13 | $Q_g$ | Gas heating value under STP conditions |
| $B$ | Dimensionless variable in equation S11 | $Q_{rt}$ | Total input heat to the reservoir from the heat source |
| $B_1$ | Dimensionless variable in equation S11 | $R$ | Universal gas constant |
| $B_a$ | Dimensionless constant in equation 10 | $R_{in}$ | Wellbore inside radius |
| BC 1 | Boundary conditions | $R_{out}$ | Wellbore outside radius |
| BC 2 | Boundary conditions | $R_w$ | Wellbore thermal resistivity |
| $b$ | Dimensionless constant in equation S14 | $r$ | Radial distance |
| $C$ | Dimensionless variable in equation S12 | $r_1$ | Wellbore's inside radius ($R_{in}$)/ Casing 1's inside radius |
| $C_{pI}$ | Zone I's specific heat capacity | $r_2$ | Casing 1's outside radius/ Gravel part's inside radius |
| $C_{pg}$ | Gas's specific heat capacity | $r_3$ | Gravel part's outside radius/ Casing 2's inside radius |
| $c$ | Dimensionless constant in equation 13 | $r_4$ | Casing 2's Outside radius/ Cement part's inside radius |
| $D$ | Dimensionless variable in equation S15 | $r_5$ | Cement part's outside radius/ Wellbore's outside radius ($R_{out}$) |
| $d$ | Dimensionless constant in equation 13 | $S$ | Interface position |
| $E$ | Dimensionless variable in equation S16 | $T_i$ | Heat source's temperature |
| $F$ | Dimensionless variable in equation S17 | $T_I$ | Temperature in Zone I |
| $F_{gH}$ | Methane gas mass ratio trapped inside the hydrate | $T_{II}$ | Temperature in Zone II |
| $G(\beta)$ | Dimensionless constant in equation S22 | $T_0$ | Initial hydrate temperature |

[Type here]

| | | | |
|---|---|---|---|
| $G_1(\beta)$ | Dimensionless constant in equation S23 | $T_s$ | Interface temperature |
| $H(\beta)$ | Dimensionless constant in equation S22 | $T_{STP}$ | Gas temperature under STP conditions |
| $H_1(\beta)$ | Dimensionless constant in equation S23 | $t$ | Time |
| $I(\beta)$ | Dimensionless constant in equation S22 | $u_r$ | Wellbore's heat flux |
| $I_1(\beta)$ | Dimensionless constant in equation S23 | $V_f$ | Volume of produced gas per moving interface's surface area in the time fraction of "$t,t-1$" |
| $K(\beta)$ | Function in equation S21 | $V_{rp}$ | Total volume of produced gas per moving interface's surface area up to time $t$ |
| $k$ | Permeability | $v_s$ | Interface velocity |
| $k_I$ | Zone I's thermal conductivity | $\rho_g$ | Gas density |
| $k_{II}$ | Zone II's thermal conductivity | $\rho_H$ | Hydrate density |
| $k_c$ | Cement's thermal conductivity | $\mu$ | Gas viscosity |
| $k_g$ | Gravel's thermal conductivity | $\lambda$ | Dimensionless variable in equation 18 |
| $k_s$ | Casing's thermal conductivity | $\beta$ | Dimensionless constant in equation 19 |
| $L(\beta)$ | Function in equation S18 | $\lambda_{os}$ | Dimensionless variable in equation 20 |
| $M(\beta)$ | Function in equation S19 | $\phi$ | Porosity |
| MH | Methane hydrate | $\alpha_I$ | Zone I's thermal diffusivity |
| $m$ | Gas molecular mass | $\alpha_{II}$ | Zone II's thermal diffusivity |
| $N(\lambda)$ | Function in equation S20 | $v_g$ | Gas velocity |
| $n_r$ | Total moles of produced gas per moving interface's surface area in the time fraction of "$t, t-1$" | $\eta_r$ | energy efficiency ratio |
| $n_{rt}$ | Total moles of produced gas per moving interface's surface area up to time $t$ | | |

# References

[1] Feng J-C, Wang Y, Li X-S, Li G, Chen Z-Y. Production behaviors and heat transfer characteristics of methane hydrate dissociation by depressurization in conjunction with warm water stimulation with dual horizontal wells. Energy. 2015;79:315-24.
[2] Davie MK, Buffett BA. A numerical model for the formation of gas hydrate below the seafloor. Journal of Geophysical Research: Solid Earth. 2001;106(B1):497-514.
[3] Collett TS. Assessment of gas hydrate resources on the North Slope. In: Conference Assessment of gas hydrate resources on the North Slope; 2008. Alaska, USA.
[4] Chong ZR, Yang SHB, Babu P, Linga P, Li X-S. Review of natural gas hydrates as an energy resource: Prospects and challenges. Applied energy. 2016;162:1633-52.
[5] Max M, Dillon WP. Oceanic methane hydrate: the character of the Blake Ridge hydrate stability zone, and the potential for methane extraction. Journal of Petroleum Geology. 1998;21(3):343-58.
[6] Englezos P. Clathrate hydrates. Industrial & engineering chemistry research. 1993;32(7):1251-74.
[7] Makogon YF. Hydrates of natural gas: PennWell Books Tulsa, OK, 1981.
[8] Collett T, Bahk J-J, Baker R, Boswell R, Divins D, Frye M, et al. Methane Hydrates in Nature-Current Knowledge and Challenges. Journal of chemical & engineering data. 2014;60(2):319-29.
[9] Li X-S, Xu C-G, Zhang Y, Ruan X-K, Li G, Wang Y. Investigation into gas production from natural gas hydrate: A review. Applied Energy. 2016;172:286-322.
[10] Li G, Li X-S, Wang Y, Zhang Y. Production behavior of methane hydrate in porous media using huff and puff method in a novel three-dimensional simulator. Energy. 2011;36(5):3170-8.
[11] Fitzgerald GC, Castaldi MJ, Zhou Y. Large scale reactor details and results for the formation and decomposition of methane hydrates via thermal stimulation dissociation. Journal of Petroleum Science and Engineering. 2012;94:19-27.
[12] Yu T, Guan G, Abudula A, Wang D. 3D visualization of fluid flow behaviors during methane hydrate extraction by hot water injection. Energy. 2019;188:116110.
[13] Yousif M, Li P, Selim M, Sloan E. Depressurization of natural gas hydrates in Berea sandstone cores. Journal of inclusion phenomena and molecular recognition in chemistry. 1990;8(1-2):71-88.
[14] Yu T, Guan G, Abudula A, Yoshida A, Wang D, Song Y. Gas recovery enhancement from methane hydrate reservoir in the Nankai Trough using vertical wells. Energy. 2019;166:834-44.
[15] Feng Y, Chen L, Suzuki A, Kogawa T, Okajima J, Komiya A, et al. Numerical analysis of gas production from layered methane hydrate reservoirs by depressurization. Energy. 2019;166:1106-19.
[16] Ji C, Ahmadi G, Smith DH. Natural gas production from hydrate decomposition by depressurization. Chemical Engineering Science. 2001;56(20):5801-14.
[17] Terzariol M, Goldsztein G, Santamarina J. Maximum recoverable gas from hydrate bearing sediments by depressurization. Energy. 2017;141:1622-8.
[18] Wang Y, Li X-S, Li G, Huang N-S, Feng J-C. Experimental study on the hydrate dissociation in porous media by five-spot thermal huff and puff method. Fuel. 2014;117:688-96.
[19] Jin G, Xu T, Xin X, Wei M, Liu C. Numerical evaluation of the methane production from unconfined gas hydrate-bearing sediment by thermal stimulation and depressurization in Shenhu area, South China Sea. Journal of Natural Gas Science and Engineering. 2016;33:497-508.
[20] Wan Q-C, Si H, Li B, Li G. Heat transfer analysis of methane hydrate dissociation by depressurization and thermal stimulation. International Journal of Heat and Mass Transfer. 2018;127:206-17.
[21] Yousif MH. Effect of under-inhibition with methanol and ethylene glycol on the hydrate control process. In: Offshore Technology Conference; 1996. Houston, Texas, USA.
[22] Sung W, Lee H, Lee H, Lee C. Numerical study for production performances of a methane hydrate reservoir stimulated by inhibitor injection. Energy Sources. 2002;24(6):499-512.

[23] Yuan Q, Sun C-Y, Yang X, Ma P-C, Ma Z-W, Liu B, et al. Recovery of methane from hydrate reservoir with gaseous carbon dioxide using a three-dimensional middle-size reactor. Energy. 2012;40(1):47-58.
[24] Ors O, Sinayuc C. An experimental study on the $CO_2$–$CH_4$ swap process between gaseous $CO_2$ and $CH_4$ hydrate in porous media. Journal of Petroleum Science and Engineering. 2014;119:156-62.
[25] Nambiar A, Babu P, Linga P. $CO_2$ capture using the clathrate hydrate process employing cellulose foam as a porous media. Canadian Journal of Chemistry. 2015;93(8):808-14.
[26] Nishikawa N, Morishita M, Uchiyama M, Yamaguchi F, Ohtsubo K, Kimuro H, et al. $CO_2$ clathrate formation and its properties in the simulated deep ocean. Energy Conversion and Management. 1992;33(5-8):651-7.
[27] Khlebnikov V, Antonov S, Mishin A, Bakulin D, Khamidullina I, Liang M, et al. A new method for the replacement of $CH_4$ with $CO_2$ in natural gas hydrate production. Natural Gas Industry B. 2016;3(5):445-51.
[28] Maruyama S, Deguchi K, Chisaki M, Okajima J, Komiya A, Shirakashi R. Proposal for a low $CO_2$ emission power generation system utilizing oceanic methane hydrate. Energy. 2012;47(1):340-7.
[29] Chen L, Sasaki H, Watanabe T, Okajima J, Komiya A, Maruyama S. Production strategy for oceanic methane hydrate extraction and power generation with Carbon Capture and Storage (CCS). Energy. 2017;126:256-72.
[30] Holder GD, Angert PF. Simulation of gas production from a reservoir containing both gas hydrates and free natural gas. In: Society of Petroleum Engineers; 1982. New Orleans, Louisiana, USA.
[31] Burshears M, O'Brien T, Malone R. A multi-phase, multi-dimensional, variable composition simulation of gas production from a conventional gas reservoir in contact with hydrates. In: Society of Petroleum Engineers; 1986. Louisville, Kentucky, USA.
[32] Yousif M, Abass H, Selim M, Sloan E. Experimental and theoretical investigation of methane-gas-hydrate dissociation in porous media. SPE reservoir Engineering. 1991;6(01):69-76.
[33] Masuda Y. Modeling and experimental studies on dissociation of methane gas hydrates in Berea sandstone cores. In: The 3rd International Conference on Gas Hydrates; 1999. Salt Lake City, Utah, USA.
[34] Masuda Y. Numerical calculation of gas production performance from reservoirs containing natural gas hydrates. In: Society of Petroleum Engineers; 1997. San Antonio, Texas, USA.
[35] Tsypkin GG. Mathematical models of gas hydrates dissociation in porous media. Annals of the New York academy of sciences. 2000;912(1):428-36.
[36] Ahmadi G, Ji C, Smith DH. Numerical solution for natural gas production from methane hydrate dissociation. Journal of petroleum science and engineering. 2004;41(4):269-85.
[37] Moridis GJ. Numerical studies of gas production from methane hydrates. In: Society of Petroleum Engineers; 2002. Calgary, Alberta, Canada.
[38] Moridis GJ, Collett TS, Dallimore SR, Satoh T, Hancock S, Weatherill B. Numerical studies of gas production from several $CH_4$ hydrate zones at the Mallik site, Mackenzie Delta, Canada. Journal of petroleum science and engineering. 2004;43(3-4):219-38.
[39] Kowalsky MB, Moridis GJ. Comparison of kinetic and equilibrium reaction models in simulating gas hydrate behavior in porous media. Energy conversion and management. 2007;48(6):1850-63.
[40] Moridis GJ, Sloan ED. Gas production potential of disperse low-saturation hydrate accumulations in oceanic sediments. Energy conversion and management. 2007;48(6):1834-49.
[41] Selim M, Sloan E. Hydrate dissociation in sediment. SPE Reservoir Engineering. 1990;5(02):245-51.
[42] McGuire PL. Recovery of gas from hydrate deposits using conventional technology. In: Society of Petroleum Engineers; 1982. Pittsburgh, Pennsylvania, USA.
[43] Makogon YF. Hydrates of hydrocarbons: Pennwell Pub. Comp Tulsa, Oklahoma. 1997.
[44] Tsypkin G. Regimes of dissociation of gas hydrates coexisting with a gas in natural strata. Journal of engineering physics and thermophysics. 2001;74(5):1083-9.
[45] Hong H, Pooladi-Darvish M, Bishnoi P. Analytical modelling of gas production from hydrates in porous media. Journal of Canadian Petroleum Technology. 2003;42(11):45-56.

[Type here]

[46] Wang Y, Feng J-C, Li X-S, Zhang Y, Li G. Analytic modeling and large-scale experimental study of mass and heat transfer during hydrate dissociation in sediment with different dissociation methods. Energy. 2015;90:1931-48.
[47] Wang Y, Feng J-C, Li X-S, Zhang Y, Li G. Large scale experimental evaluation to methane hydrate dissociation below quadruple point in sandy sediment. Applied energy. 2016;162:372-81.
[48] Tang L-G, Li X-S, Feng Z-P, Li G, Fan S-S. Control mechanisms for gas hydrate production by depressurization in different scale hydrate reservoirs. Energy & Fuels. 2007;21(1):227-33.
[49] Li X-S, Wang Y, Li G, Zhang Y, Chen Z-Y. Experimental investigation into methane hydrate decomposition during three-dimensional thermal huff and puff. Energy & Fuels. 2011;25(4):1650-8.
[50] Li X-S, Yang B, Zhang Y, Li G, Duan L-P, Wang Y, et al. Experimental investigation into gas production from methane hydrate in sediment by depressurization in a novel pilot-scale hydrate simulator. Applied energy. 2012;93:722-32.
[51] Zhao J, Cheng C, Song Y, Liu W, Liu Y, Xue K, et al. Heat transfer analysis of methane hydrate sediment dissociation in a closed reactor by a thermal method. Energies. 2012;5(5):1292-308.
[52] Wang Y, Feng J-C, Li X-S, Zhang Y, Chen Z-Y. Fluid flow mechanisms and heat transfer characteristics of gas recovery from gas-saturated and water-saturated hydrate reservoirs. International Journal of Heat and Mass Transfer. 2018;118:1115-27.
[53] Wang X, Dong B, Wang F, Li W, Song Y. Pore-scale investigations on the effects of ice formation/melting on methane hydrate dissociation using depressurization. International Journal of Heat and Mass Transfer. 2019;131:737-49.
[54] Konno Y, Fujii T, Sato A, Akamine K, Naiki M, Masuda Y, et al. Key findings of the world's first offshore methane hydrate production test off the coast of Japan: Toward future commercial production. Energy & Fuels. 2017;31(3):2607-16.
[55] Chen W, Hartman RL. Methane Hydrate Intrinsic Dissociation Kinetics Measured in a Microfluidic System by Means of in Situ Raman Spectroscopy. Energy & Fuels. 2018;32(11):11761-71.
[56] Mardani M, Azimi A, Javanmardi J, Mohammadi AH. Effect of EMIM-BF4 Ionic Liquid on Dissociation Temperature of Methane Hydrate in the Presence of PVCap: Experimental and Modeling Studies. Energy & Fuels. 2018;33(1):50-7.
[57] Wang Y, Feng J-C, Li X-S, Zhang Y. Experimental and modeling analyses of scaling criteria for methane hydrate dissociation in sediment by depressurization. Applied energy. 2016;181:299-309.
[58] Zhao J, Fan Z, Dong H, Yang Z, Song Y. Influence of reservoir permeability on methane hydrate dissociation by depressurization. International Journal of Heat and Mass Transfer. 2016;103:265-76.
[59] Ding Y-L, Xu C-G, Yu Y-S, Li X-S. Methane recovery from natural gas hydrate with simulated IGCC syngas. Energy. 2017;120:192-8.
[60] Roostaie M, Leonenko Y. Analytical modeling of methane hydrate dissociation under thermal stimulation. Journal of Petroleum Science and Engineering. 2019:106505.
[61] Wang H, Li X, Sepehrnoori K, Zheng Y, Yan W. Calculation of the wellbore temperature and pressure distribution during supercritical CO2 fracturing flowback process. International Journal of Heat and Mass Transfer. 2019;139:10-6.
[62] Xiong W, Bahonar M, Chen Z. Development of a thermal wellbore simulator with focus on improving heat loss calculations for SAGD steam injection. In: Society of Petroleum Engineers; 2015. Calgary, Alberta, Canada.
[63] Sun F, Yao Y, Li X, Li G, Sun Z. A numerical model for predicting distributions of pressure and temperature of superheated steam in multi-point injection horizontal wells. International Journal of Heat and Mass Transfer. 2018;121:282-9.
[64] Florez Anaya A, Osorio MA. A Successful Gravel-Packing Technique in Vertical and Deviated Wells with Enlarged Open Hole in Cased Completions: A Case Study, Rubiales and Quifa Fields. In: Society of Petroleum Engineers; 2014. Medellín, Colombia.
[65] Pucknell J, Mason J. Predicting the Pressure Drop in a Cased-Hole Gravel Pack Completion. In: Society of Petroleum Engineers; 1992. Cannes, France.

[66] Wang L, Liu H, Pang Z, Lv X. Overall heat transfer coefficient with considering thermal contact resistance in thermal recovery wells. International Journal of Heat and Mass Transfer. 2016;103:486-500.
[67] Xu E, Soga K, Zhou M, Uchida S, Yamamoto K. Numerical analysis of wellbore behaviour during methane gas recovery from hydrate bearing sediments. In: Offshore Technology Conference; 2014. Houston, Texas, USA.
[68] Tsimpanogiannis IN, Lichtner PC. Parametric study of methane hydrate dissociation in oceanic sediments driven by thermal stimulation. Journal of Petroleum Science and Engineering. 2007;56(1-3):165-75.
[69] Weinbaum S, Wheeler Jr H. Heat Transfer in Sweat-Cooled Porous Metals. Journal of Applied Physics. 1949;20(1):113-22.
[70] Carslaw H, Jaeger J. Conduction of heat in solids: Oxford Science Publications: Oxford, England, 1959.
[71] Özışık MN. Heat conduction: John Wiley & Sons, 1993.
[72] Song Y, Cheng C, Zhao J, Zhu Z, Liu W, Yang M, et al. Evaluation of gas production from methane hydrates using depressurization, thermal stimulation and combined methods. Applied Energy. 2015;145:265-77.
[73] Remund CP. Borehole thermal resistance: laboratory and field studies. ASHRAE transactions. 1999;105:439.
[74] Dalla Santa G, Peron F, Galgaro A, Cultrera M, Bertermann D, Mueller J, et al. Laboratory measurements of gravel thermal conductivity: An update methodological approach. Energy Procedia. 2017;125:671-7.
[75] Cheng W-L, Huang Y-H, Lu D-T, Yin H-R. A novel analytical transient heat-conduction time function for heat transfer in steam injection wells considering the wellbore heat capacity. Energy. 2011;36(7):4080-8.
[76] Liang H, Song Y, Chen Y. Numerical simulation for laboratory-scale methane hydrate dissociation by depressurization. Energy Conversion and Management. 2010;51(10):1883-90.
[77] Li G, Li X-S, Li B, Wang Y. Methane hydrate dissociation using inverted five-spot water flooding method in cubic hydrate simulator. Energy. 2014;64:298-306.
[78] Li X-S, Wang Y, Li G, Zhang Y. Experimental investigations into gas production behaviors from methane hydrate with different methods in a cubic hydrate simulator. Energy & Fuels. 2011;26(2):1124-34.
[79] Wang Y, Li X-S, Li G, Zhang Y, Li B, Chen Z-Y. Experimental investigation into methane hydrate production during three-dimensional thermal stimulation with five-spot well system. Applied energy. 2013;110:90-7.
[80] Tang LG, Xiao R, Huang C, Feng Z, Fan SS. Experimental investigation of production behavior of gas hydrate under thermal stimulation in unconsolidated sediment. Energy & Fuels. 2005;19(6):2402-7.
[81] Bayles G, Sawyer W, Anada H, Reddy S, Malone R. A steam cycling model for gas production from a hydrate reservoir. Chemical Engineering Communications. 1986;47(4-6):225-45.
[82] Zhao J, Wang J, Liu W, Song Y. Analysis of heat transfer effects on gas production from methane hydrate by thermal stimulation. International Journal of Heat and Mass Transfer. 2015;87:145-50.
[83] Zhao J, Liu D, Yang M, Song Y. Analysis of heat transfer effects on gas production from methane hydrate by depressurization. International Journal of Heat and Mass Transfer. 2014;77:529-41.

[Type here]